\documentclass[preprint,preprintnumbers,amsmath,amssymb,floatfix,nofootinbib]{revtex4}
\usepackage{epsf, array, color}
\usepackage{graphicx}
\usepackage{subfigure}
\usepackage{bbold}
\usepackage{amsmath}
\usepackage{mathtools}
\usepackage{color}
\usepackage{dcolumn}% Align table columns on decimal point
\usepackage{hyperref}

\newcommand{\bea}{\begin{eqnarray}}
\newcommand{\eea}{\end{eqnarray}}

\def\beq{\begin{equation}}
\def\eeq{\end{equation}}

\begin{document}

%\preprint{}

\title{Benchmarks for Double Higgs Production in the Singlet Extended Standard Model at the LHC}

\affiliation{Department of Physics and Astronomy, University of Kansas, Lawrence, Kansas, 66045 USA}
\author{Ian~M.~Lewis and Matthew~Sullivan}\affiliation{Department of Physics and Astronomy, University of Kansas, Lawrence, Kansas, 66045 USA}

%\date{\today}

\begin{abstract}
The simplest extension of the Standard Model is to add a gauge singlet scalar, $S$: the singlet extended Standard Model.  In the absence of a $Z_2$ symmetry $S\rightarrow -S$ and if the new scalar is sufficiently heavy, this model can lead to resonant double Higgs production, significantly increasing the production rate over the Standard Model prediction.   While searches for this signal are being performed, it is important to have benchmark points and models with which to compare the experimental results.  In this paper we determine these benchmarks by maximizing the double Higgs production rate at the LHC in the singlet extended Standard Model.  We find that, within current constraints, the branching ratio of the new scalar into two Standard Model-like Higgs bosons can be upwards of $0.76$, and the double Higgs rate can be increased upwards of 30 times the Standard Model prediction.

\end{abstract}

\maketitle

\section{\label{sec:intro} Introduction}
One of the main objectives of the Large Hadron Collider (LHC) is to further our understanding of electroweak (EW) physics at the EW scale.  Of particular interest are the interactions of the observed Higgs boson~\cite{Chatrchyan:2012xdj,Aad:2012tfa}.  In fact, measurements of the Higgs production and decay rates are at the level of $\sim 20\%$ precision~\cite{Khachatryan:2016vau}. Although these measurements help us determine if the observed Higgs boson is related to the source of fundamental masses within the Standard Model (SM), there are still many unanswered questions.  One of the most pressing is the mechanism of EW symmetry breaking (EWSB).  In the SM the source of EWSB is the scalar potential.  Hence, it is interesting to study extensions of the SM that change the potential and their signatures at the LHC.  In particular, simple extensions allow us to investigate phenomenology that is generic to more complete models.

The simplest extension of the SM is the addition of a gauge singlet real scalar, $S$: the singlet extended SM.  At the renormalizable level, the only allowed interactions between $S$ and the SM are with the Higgs field.  Hence, this model is a useful laboratory to investigate deviations from the SM Higgs potential.  Although this is the simplest possible extension, it is well-motivated.  This scenario arises in Higgs portal models~\cite{Binoth:1996au,Davoudiasl:2004be,Schabinger:2005ei,Patt:2006fw,Bowen:2007ia,Bock:2010nz,Djouadi:2011aa,Englert:2011aa,Englert:2011yb,Alanne:2014bra}.  In these models, the scalar singlet couples to a dark matter sector.  Through its interactions with the Higgs field, the new scalar provides couplings between the dark sector and the SM.  Additionally, scalar singlets can help provide the strong first order EW phase transition necessary for EW baryogenesis~\cite{Ham:2004cf,Profumo:2007wc,Ashoorioon:2009nf,Bodeker:2009qy,Espinosa:2011ax,No:2013wsa,Alanne:2014bra,Curtin:2014jma,Profumo:2014opa,Huang:2015bta,Damgaard:2015con,Kozaczuk:2015owa,Huang:2015tdv,Xiao:2015tja,
Huang:2016cjm,Kotwal:2016tex,Curtin:2016urg,Vaskonen:2016yiu}.

If there is no $Z_2$ symmetry, $S\rightarrow -S$, after EWSB the new scalar will mix with the SM Higgs boson.  This mixing induces couplings between the new scalar and the rest of the SM particles.  Hence, the new scalar can be produced and searched for at the LHC, as well as affecting precision Higgs measurements.  The simplicity of the singlet extended SM allows for easy interpretation of precision Higgs measurements~\cite{Aad:2015pla,Khachatryan:2016vau} and resonant searches for heavy scalars~\cite{Aad:2015agg,Khachatryan:2015cwa,Aad:2015kna,Aad:2015xja,Khachatryan:2015yea,Khachatryan:2016sey,CMS:2016knm,CMS:2016tlj,ATLAS:2016ixk,ATLASbbgamgam,CMS:2016ilx,CMS:2016jpd,ATLAS:2016oum,ATLAS:2016kjy,Aaboud:2016okv,CMS:2016zte,
Khachatryan:2015sma,Aad:2015fna}.  

There have been many phenomenological studies of the singlet extended SM at the LHC~\cite{Schabinger:2005ei,OConnell:2006rsp,Bowen:2007ia,Profumo:2007wc,Barger:2008jx,Dawson:2009yx,Englert:2011yb,Bertolini:2012gu,Pruna:2013bma,Caillol:2013gqa,Curtin:2014jma,Lopez-Val:2014jva,Profumo:2014opa,Buttazzo:2015bka,Robens:2015gla,Falkowski:2015iwa,Bojarski:2015kra,Costa:2015llh,Fischer:2016rsh,Fichet:2016xpw}.  
Of particular interest to us is if the new scalar is sufficiently heavy, it can decay on-shell into two SM-like Higgs bosons, mediating resonant double Higgs production at the LHC~\cite{Dolan:2012ac,Cooper:2013kia,No:2013wsa,Chen:2014ask,Martin-Lozano:2015dja,Dawson:2015haa,Godunov:2015nea,Robens:2016xkb,Nakamura:2017irk,Huang:2017jws,Chang:2017niy,Lu:2015qqa,Ren:2017jbg}.  This can greatly enhance the double Higgs rate over the SM prediction.  We will provide benchmark points that maximize double Higgs production in the singlet extended SM.  These benchmark points are needed to help determine when the experimental searches for resonant double Higgs production~\cite{ATLAS:2016ixk,ATLASbbgamgam,CMS:2016tlj,CMS:2016knm,Khachatryan:2015yea,Khachatryan:2016sey,Aad:2015xja} are probing interesting regions of parameter space.\footnote{A similar study has been done in the case of a broken $Z_2$ symmetry $S\rightarrow -S$~\cite{Robens:2016xkb}.  Here we work in the singlet extended SM with no $Z_2$.  This model has more free parameters allowing for different benchmark rates.}

In Section~\ref{sec:model} we provide an overview of the model, including the theoretical constraints on the model.  Experimental constraints are discussed in Section~\ref{sec:constraint}.  Resonant double Higgs production is discussed in Section~\ref{sec:DoubleHiggs}.  In Section~\ref{sec:results} we discuss the maximization of the double Higgs rate and provide the benchmark points.  We conclude with Section~\ref{sec:conc}.

%%%%%%%%%%%%%%%%%%%%%%%%%%%%%%%%%%%%%%%%%%%%%%%%%%%%%%%%%%%%%
\section{\label{sec:model} The Singlet Extended Standard Model}
%%%%%%%%%%%%%%%%%%%%%%%%%%%%%%%%%%%%%%%%%%%%%%%%%%%%%%%%%%%%%
In this section we give an overview of the singlet extended SM, following the notation of Ref.~\cite{Chen:2014ask}. The results of Ref.~\cite{Chen:2014ask} are important for establishing our benchmark points. Hence, we summarize the results of this paper regarding global minimization of the potential, vacuum stability, and perturbative unitarity. In the remaining part of the paper we will extend upon this work, thoroughly investigating the relationship of these theoretical constraints and maximization of double Higgs production.

The model contains the SM Higgs doublet, $H$, and a new real gauge singlet scalar, $S$. The new singlet does not directly couple to SM particles except for the Higgs doublet. Allowing for all renormalizable terms, the most general scalar potential is
\begin{equation}
V(H,S) = - \mu^2 H^\dagger H + \lambda (H^\dagger H)^2 + \frac{a_1}{2} H^\dagger H S + \frac{a_2}{2} H^\dagger H S^2 +b_1 S + \frac{b_2}{2} S^2 + \frac{b_3}{3} S^3 + \frac{b_4}{4} S^4. \label{potential}
\end{equation}
The neutral scalar component of $H$ is denoted as $\phi_0 = (h+v)/\sqrt{2}$ with the vacuum expectation value (vev) being $\langle \phi_0 \rangle = \frac{v}{\sqrt{2}}$. We similarly write $S=s+x$, where the vev of $S$ is denoted as $x$.

We require that EWSB occurs at an extremum of the potential, so that $v=v_{EW}=246$ GeV. Shifting the field $S\rightarrow S+\delta S$ does not introduce any new terms to the potential, and is only a meaningless change in parameters. Using this freedom, we can additionally choose that the EWSB minimum satisfies $x=0$.  Requiring that $(v,x)=(v_{EW},0)$ be an extremum of the potential gives
\begin{eqnarray} 
\mu^2 &=& \lambda v_{EW}^2,\nonumber\\
b_1 &=& - \frac{v_{EW}^2}{4} a_1.
\end{eqnarray}

After symmetry breaking, there are two mass eigenstates denoted as $h_1$ and $h_2$ with masses $m_1$ and $m_2$, respectively. The new fields are related to the gauge eigenstate fields by
\begin{equation}
\begin{pmatrix}
h_1 \\ h_2
\end{pmatrix} = 
\begin{pmatrix}
\cos \theta & \sin \theta \\
-\sin \theta & \cos \theta
\end{pmatrix}
\begin{pmatrix}
h \\ s
\end{pmatrix}.
\end{equation}
where $\theta$ is the mixing angle. The masses, $m_1$ and $m_2$, and the mixing angle, $\theta$, are related to the scalar potential parameters
\begin{eqnarray}
a_1 &=& \frac{m_1^2 - m_2^2}{v_{EW}} \sin{2 \theta},\nonumber\\
b_2 + \frac{a_2}{2} v_{EW}^2 &=& m_1^2 \sin^2{\theta} + m_2^2 \cos^2{\theta},\nonumber\\
\lambda &=& \frac{ m_1^2 \cos^2{\theta} + m_2^2 \sin^2{\theta}}{2 v_{EW}^2}.
\end{eqnarray}
We set the mass $m_1=125$ GeV to reproduce the discovered Higgs. The free parameter space is then 
\begin{equation}
m_2\textrm{, }\theta \textrm{, }a_2\textrm{, }b_3\textrm{, and  }b_4.
\end{equation}

We are interested in the scenario with $m_2 \geq 2 m_1$, where $h_2$ can decay on-shell to two SM-like Higgs bosons, $h_1$. After symmetry breaking, the trilinear scalar terms in the potential which are relevant to double Higgs production are

\begin{equation}
V(h_1, h_2) \supset \frac{\lambda_{111}}{3!} h_1^3  + \frac{\lambda_{211}}{2!} h_2 h_1^2. \label{trilinears}
 \end{equation}
The trilinear coupling $\lambda_{211}$ allows for the tree level decay of $h_2 \rightarrow h_1 h_1 $. At the EWSB minimum $(v,x)=(v_{EW},0)$,  the trilinear couplings are given by~\cite{Chen:2014ask}
 \begin{eqnarray}
\lambda_{111} &=& 2 \sin^3\theta \,b_3 + \frac{3 a_1}{2} \sin\theta \cos^2\theta + 3\, a_2 \sin^2\theta \cos\theta\, v_{EW} + 6 \,\lambda\,\cos^3\theta\, v_{EW},\label{eq:trilinear}\\
\lambda_{211} &=& 2 \sin^2\theta \cos\theta\, b_3 + \frac{a_1}{2} \cos\theta (\cos^2\theta - 2 \sin^2\theta) + (2 \cos^2\theta - \sin^2\theta) \sin\theta\, v_{EW}\, a_2\nonumber\\
&& - 6 \lambda \sin\theta \cos^2\theta\, v_{EW}.\nonumber
\end{eqnarray}
%%%%%%%%%%%%%%%%%%%%%%%%%%%%%%%%%%%%%%%%%%%%%%%%%%%%%%%%%%%%%%%
\subsection{\label{sec:minimization} Global Minimization of the Potential}
%%%%%%%%%%%%%%%%%%%%%%%%%%%%%%%%%%%%%%%%%%%%%%%%%%%%%%%%%%%%%%%

The scalar potential, Eq~(\ref{potential}), allows for many extrema $(v,x)$.  There are two classes that need to be considered: $v\neq0$ and $v = 0$. The $v \neq 0$ extrema are given by $(v,x)=(v_{EW},0)$ and  $(v,x)=(v_\pm,x_\pm)$ where~\cite{Chen:2014ask}
\begin{eqnarray}
x_\pm&\equiv& \frac{v_{EW}(3a_1 a_2-8 b_3\lambda)\pm 8\sqrt{\Delta}}{4v_{EW}(4b_4\lambda-a_2^2)},\nonumber\\
v^2_\pm&\equiv& v_{EW}^2-\frac{1}{2\lambda}\left(a_1x_\pm + a_2 x^2_\pm\right),\nonumber\\
\Delta &=& \frac{v^2_{EW}}{64}\left(8b_3\lambda-3a_1a_2\right)^2-\frac{m_1^2m_2^2}{2}\left(4b_4\lambda-a_2^2\right).
\end{eqnarray}
For all of these three solutions to be real, there are constraints $\Delta>0$ and $v_{\pm}^2>0$.

The $v = 0$ extrema are given by solutions of the following cubic equation:
\begin{equation}
b_1 + b_2 x + b_3 x^2 + b_4 x^3 = 0. \label{cubic}
\end{equation}
Only real solutions for $x$ are of interest. Manifestly real solutions for non-degenerate cubics are presented in the Appendix.

As can be seen, there is only one extremum with $v=v_{EW}$. Since the scalar $S$ is a gauge singlet, it does not contribute to the gauge boson or SM fermion masses.  Hence, to reproduce the correct EWSB pattern, we require that $(v_{EW},0)$ is the global minimum.
%%%%%%%%%%%%%%%%%%%%%%%%%%%%%%%%%%%%%%%%%%%%%%%%%%%%%%%%%%%%%%%
\subsection{\label{sec:positivedefinite} Vacuum Stability}
%%%%%%%%%%%%%%%%%%%%%%%%%%%%%%%%%%%%%%%%%%%%%%%%%%%%%%%%%%%%%%%
To avoid instability of the vacuum from runaway negative energy solutions, the scalar potential should be bounded from below at large field values. Vacuum stability of the potential then requires that
\begin{equation}
4 \lambda \phi_0^4  + 2 a_2 \phi_0^2 s^2 + b_4 s^2 > 0. \label{posdef}
\end{equation}
It is clear that bounding the potential from below along the axes $s=0$ and $\phi_0$=0 requires 
\begin{eqnarray}
\lambda > 0\quad\quad{\rm and}\quad\quad b_4 > 0.
\end{eqnarray}
 If $a_2 > 0$ as well, then the potential is always positive definite for large field values. However, $a_2 < 0$ is also allowed. Eq.~(\ref{posdef}) can be rewritten as
\begin{equation}
\lambda (2 \phi_0^2 + \frac{a_2}{2 \lambda} s^2)^2 + (b_4 - \frac{a_2^2}{4 \lambda}) s^4 > 0. \label{posdefrewrite}
\end{equation}
The first term in Eq.~(\ref{posdefrewrite}) is always positive definite.  Requiring the second term to be nonnegative for $a_2<0$ gives the bound~\cite{Chen:2014ask}
\begin {equation}
-2 \sqrt{\lambda b_4} \leq a_2.
\end{equation}
%%%%%%%%%%%%%%%%%%%%%%%%%%%%%%%%%%%%%%%%%%%%%%%%%%%%%%%%%%%%%%%
\subsection{\label{sec:unitarity} Perturbative Unitarity}
%%%%%%%%%%%%%%%%%%%%%%%%%%%%%%%%%%%%%%%%%%%%%%%%%%%%%%%%%%%%%%%
Perturbative unitarity of the partial wave expansion for the scattering also constrains quartic scalar couplings,
\begin{eqnarray}
\mathcal{M}=16\pi\sum_{j=0}^{\infty} (2j+1)a_j P_j(\cos\theta),
\end{eqnarray}
where $P_j(\cos\theta)$ are Legendre polynomials.
 Looking at the process $h_2 h_2 \rightarrow h_2 h_2$ for large energies, the first term in the partial wave expansion at leading order is
\begin{equation}
a_0(h_2 h_2 \rightarrow h_2 h_2) = \frac{3 b_4} {8 \pi}.
\end{equation}
The perturbative unitarity requirement $|a_0| \leq 0.5$ gives the constraint $b_4 \lesssim 4.2$.  When this bound is saturated, a minimum higher order correction of $41\%$ is needed to restore the unitarity of the amplitude~\cite{Schuessler:2007av}.

There are also perturbative unitarity constraints on the other quartic couplings: $\lambda\lesssim4.2$ and $a_2\lesssim25$.  However, for all parameter points we consider, these constraints on $\lambda$ and $a_2$ are automatically satisfied when all other constraints are applied.
%%%%%%%%%%%%%%%%%%%%%%%%%%%%%%%%%%%%%%%%%%%%%%%%%%%%%%%%%%%%%
\section{\label{sec:constraint}Experimental Constraints}
%%%%%%%%%%%%%%%%%%%%%%%%%%%%%%%%%%%%%%%%%%%%%%%%%%%%%%%%%%%%%

The singlet model predicts that the couplings of $h_1$ to other SM fermions and gauge bosons are suppressed from the SM predictions by $\cos \theta$.  Hence, the single Higgs production cross section is suppressed by $\cos^2 \theta$,
\begin{equation}
\sigma(pp \rightarrow h_1) = \cos^2 \theta \sigma_{SM}(pp \rightarrow h_1)\label{eq:h1prod}
\end{equation}
where $\sigma_{SM}(pp \rightarrow h_1)$ is the SM cross section for Higgs production at $m_1=125$~GeV.  Since all couplings between $h_1$ and SM fermions and gauge bosons are universally suppressed, the branching ratios for $h_1$ decay agree with SM branching ratios,
\begin{equation}
\textrm{BR}(h_1 \rightarrow X_{SM}) =\textrm{BR}_{SM}(h_1 \rightarrow X_{SM})\label{eq:h1dec}
\end{equation}
where $X_{SM}$ is any allowed SM final state.
Using these properties, the most stringent constraint from observed Higgs signal strengths is from ATLAS: $\sin^2{\theta} \leq 0.12$ at $95\%$ C.~L.~\cite{Aad:2015pla}. 

As mentioned earlier, there are also direct constraints from searches for heavy scalar particles~\cite{Khachatryan:2015cwa,Aad:2015kna,Aad:2015agg,Aad:2015xja,Khachatryan:2016sey,Khachatryan:2015yea,CMS:2016knm,CMS:2016tlj,ATLAS:2016ixk,ATLASbbgamgam,CMS:2016ilx,CMS:2016jpd,ATLAS:2016oum,ATLAS:2016kjy,Aaboud:2016okv,CMS:2016zte,
Khachatryan:2015sma,Aad:2015fna}. For the mass range $250 \textrm{ GeV} \leq m_2 \leq 1000 \textrm{ GeV}$ considered here, the direct constraints on $\sin\theta$ are weaker than those from the Higgs signal strengths~\cite{Robens:2016xkb}.  Nevertheless, independently and using {\tt HiggsBounds}~\cite{arXiv:0811.4169, arXiv:1102.1898, arXiv:1301.2345, arXiv:1311.0055, arXiv:1507.06706}, we verify that our benchmark points satisfy all experimental constraints.
%%%%%%%%%%%%%%%%%%%%%%%%%%%%%%%%%%%%%%%%%%%%%%%%%%%%%%%%%%%%%
\section{\label{sec:DoubleHiggs} Production and Decay Rates}
%%%%%%%%%%%%%%%%%%%%%%%%%%%%%%%%%%%%%%%%%%%%%%%%%%%%%%%%%%%%%

\begin{figure}[tb]
\begin{center}
\subfigure[]{\includegraphics[width=0.48\textwidth,clip]{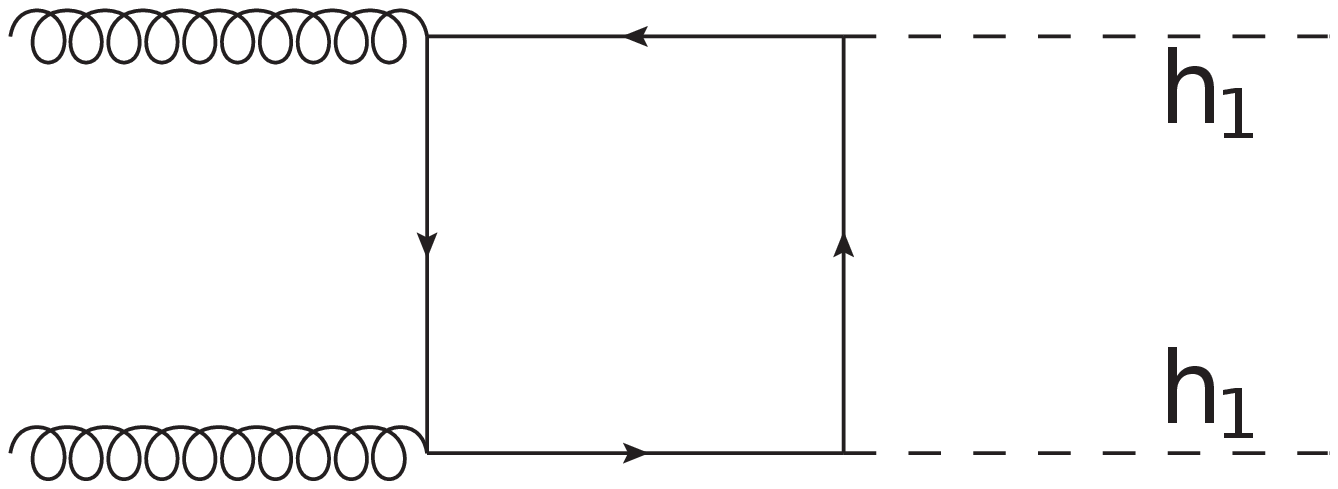}\label{fig:box}}
\subfigure[]{\includegraphics[width=0.48\textwidth,clip]{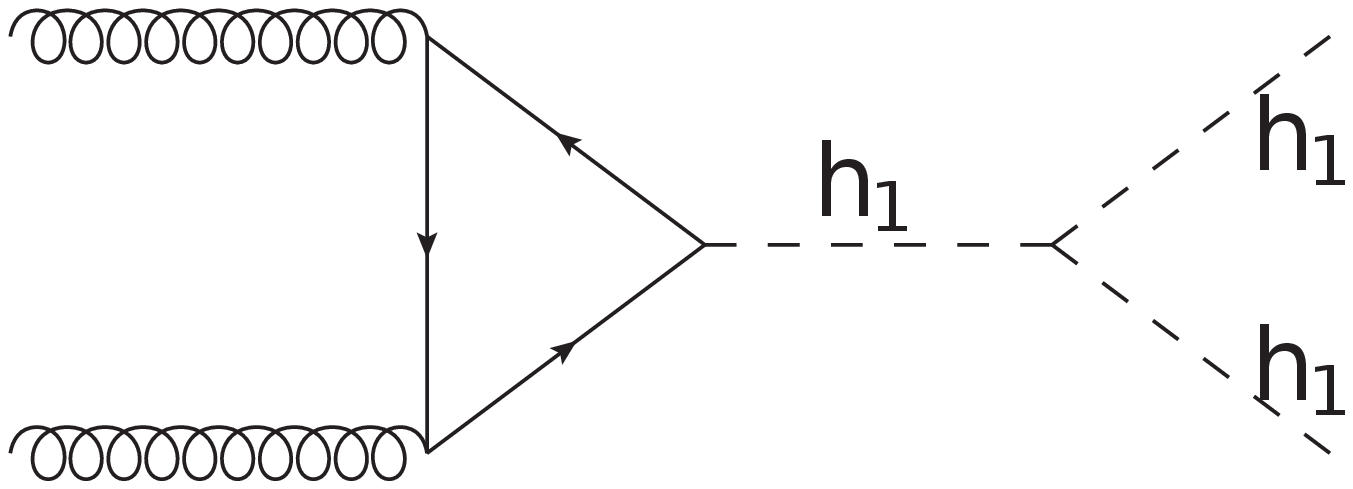}\label{fig:tri_h1}}\\
\subfigure[]{\includegraphics[width=0.48\textwidth,clip]{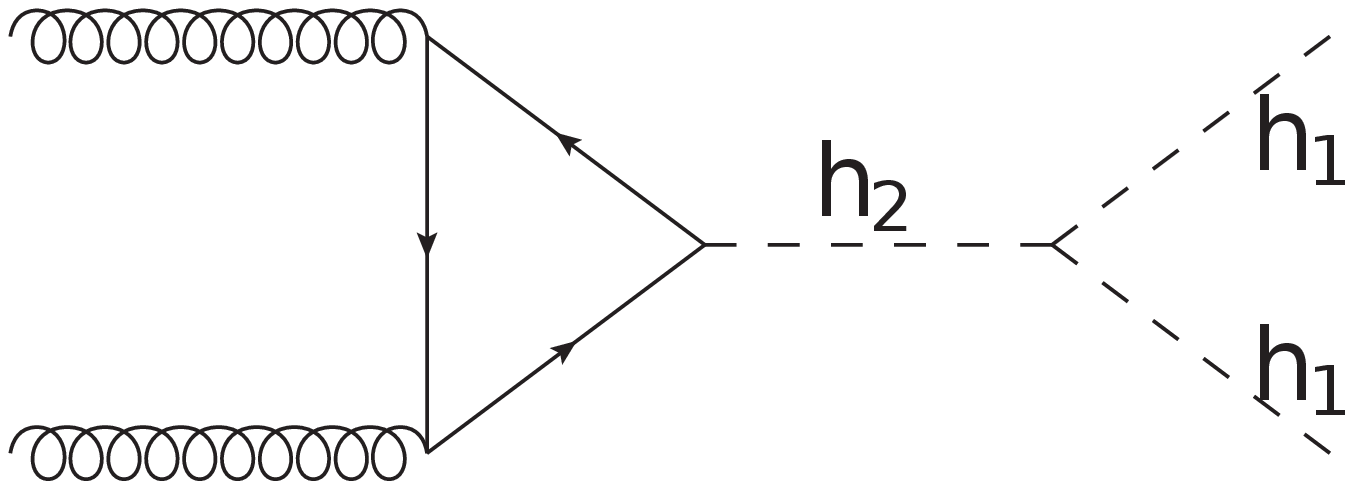}\label{fig:tri_h2}}
\end{center}
\caption{Representative diagrams for  double Higgs production corresponding to (a) box diagram, (b) triangle diagram with the $s$-channel SM-like Higgs boson $h_1$, 
and (c) triangle diagram with the resonant $s$-channel $h_2$. The top quark loops are the dominant contribution to the production.}
\label{fig:gghh}
\end{figure}

The contributions to double Higgs production in the singlet model are shown in Fig.~\ref{fig:gghh}.   Figures~\ref{fig:box} and~\ref{fig:tri_h1} are present in the SM double Higgs production, while the $s$-channel $h_2$ contribution in Fig.~\ref{fig:tri_h2} is responsible for the resonant $h_1h_1$ production.  The $s$-channel $h_1$ ($h_2$) contribution in Fig.~\ref{fig:tri_h1} (Fig.~\ref{fig:tri_h2}) depends on the scalar trilinear couplings $\lambda_{111}$ ($\lambda_{211}$) in Eq.~\ref{eq:trilinear}.  Hence, this process is clearly sensitive to the shape of the scalar potential.

It is expected that the resonant $h_2$ contribution dominates the double Higgs production cross section.  We then use the narrow width approximation as follows:
\begin{equation}
\sigma(pp  \rightarrow h_2 \rightarrow  h_1 h_1) \approx \sigma(pp  \rightarrow h_2) \textrm{BR}(h_2 \rightarrow h_1 h_1)\textrm{.}\label{eq:res}
\end{equation}
Although interference effects between the different contributions in Fig.~\ref{fig:gghh} can be significant~\cite{Dawson:2015haa}, our purpose here is to maximize the double Higgs rate in this model.  Hence, for simplicity we focus on maximizing the cross section in Eq.~(\ref{eq:res}).  This is sufficient to attain our goal. 

Due to mixing with the Higgs boson, $h_2$ has couplings to SM fermions and gauge bosons proportional to $\sin{\theta}$. The cross section for production of $h_2$ is then
\begin{equation}
\sigma(pp \rightarrow h_2) = \sin^2 \theta\, \sigma_{SM}(pp \rightarrow h_2)\label{eq:h2prod}
\end{equation}
with $\sigma_{SM}(pp \rightarrow h_2)$ being the SM Higgs production cross section evaluated at a Higgs mass of $m_2$.  Since the couplings to fermions and gauge bosons are proportional to the SM values, the intuition about the dominant SM Higgs production channels is valid for the production of $h_2$.  Hence, gluon fusion $gg\rightarrow h_2$ is the dominant channel, as illustrated in Fig.~\ref{fig:tri_h2}.

The heavy scalar $h_2$ can decay to SM gauge bosons and fermions with partial widths of
\begin{equation}
\Gamma(h_2 \rightarrow X_{SM}) = \sin^2 \theta\, \Gamma_{SM}(h_2\rightarrow X_{SM})\label{eq:h2SMdec}
\end{equation}
where $\Gamma_{SM}(h_2\rightarrow X_{SM})$ is the SM decay width for a Higgs boson into SM final states $X_{SM}\neq h_1h_1$ evaluated at a mass of $m_2$.  The tree level decay for $h_2 \rightarrow h_1 h_1$ has a partial width given by

\begin{equation}
\Gamma(h_2 \rightarrow h_1 h_1) = \frac{\lambda_{211}^2}{32 \pi m_2} \sqrt{1 - \frac{4 m_1^2}{m_2^2}}
\end{equation}
The branching ratio for $h_2 \rightarrow h_1 h_1$ is

\begin{equation}
\textrm{BR}(h_2 \rightarrow h_1 h_1) = \frac{ \Gamma(h_2 \rightarrow h_1 h_1)}{\Gamma(h_2)},
\end{equation}
where 
\begin{eqnarray}
\Gamma(h_2)=\Gamma(h_2 \rightarrow h_1 h_1) + \sin^2 \theta \,\Gamma_{SM}(h_2\rightarrow X_{SM})
\end{eqnarray}
is the total width of $h_2$.

\begin{figure}[t]
\subfigure[]{\includegraphics[width=0.45\textwidth,clip]{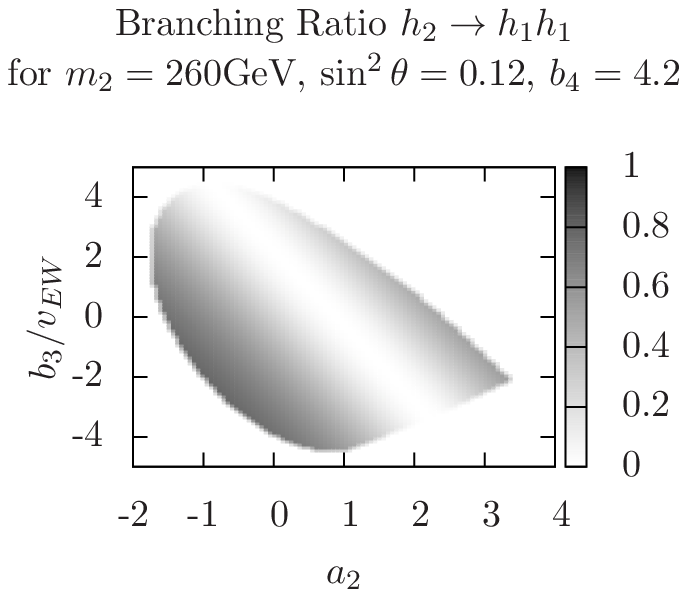}}\hspace{0.3in}
\subfigure[]{\includegraphics[width=0.45\textwidth,clip]{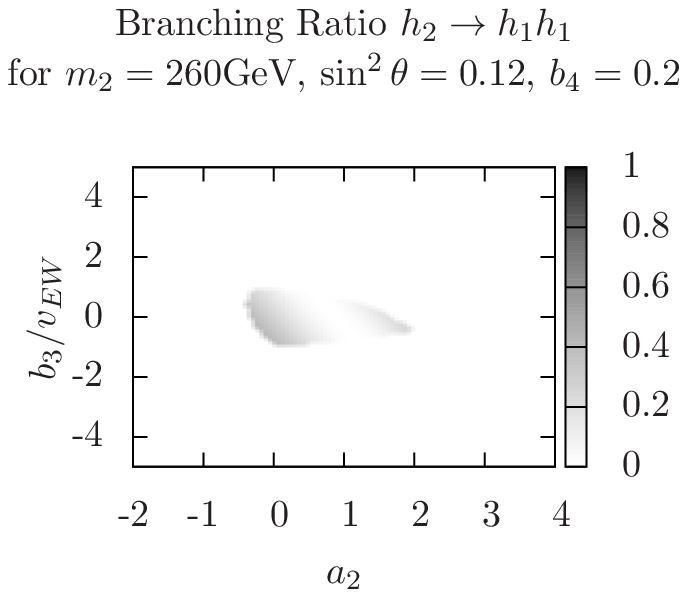}}
\caption{\label{fig:paramlargeb4}$\textrm{BR}(h_2 \rightarrow h_1 h_1)$ as a function of $b_3$ and $a_2$ for $m_2=260 \textrm{ GeV}$ and  $\sin^2{\theta}=0.12$.  In (a) $b_4=4.2$ and (b) $b_4=0.2$. The shaded regions are allowed by the global minimum constraint.  The darker shaded regions have larger $\textrm{BR}(h_2 \rightarrow h_1 h_1)$.}
\end{figure}

 The parameter $b_4$ does not explicitly affect $\textrm{BR}(h_2 \rightarrow h_1 h_1)$.  However, through the constraints of vacuum stability and $(v,x)=(v_{EW},0)$ being the global minimum of the scalar potential [Sec.~\ref{sec:minimization}], $b_4$ affects the allowed ranges for the other parameters $a_2$ and $b_3$.  These parameters appear in the trilinear coupling $\lambda_{211}$ in Eq.~(\ref{eq:trilinear}), which is relevant for $\Gamma(h_2\rightarrow h_1h_1)$.   
Figure~\ref{fig:paramlargeb4} shows the allowed parameter region satisfying these constraints for (a) $b_4=4.2$ and (b) $b_4=0.2$ with $m_2=260$~GeV and $\sin^2\theta=0.12$. It is clear from the figures that a lower value of $b_4$ shrinks the allowed region. The shading in the figures indicates the value of $\textrm{BR}(h_2 \rightarrow h_1 h_1)$, where the values of $\Gamma_{SM}(h_2\rightarrow X_{SM})$ were obtained from Ref.~\cite{deFlorian:2016spz}.
It was found that the maximum ${\rm BR}(h_2\rightarrow h_1h_1)$ always occurs with $b_4=4.2$ at the unitarity bound.

\begin{figure}[tb]
\includegraphics[width=0.48\textwidth,clip]{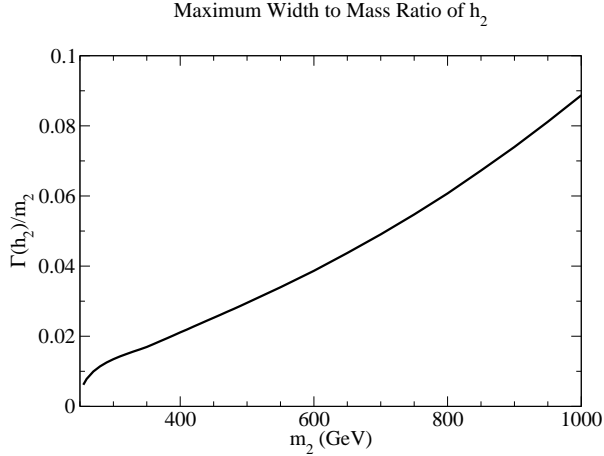}
\caption{\label{fig:totwidth} The ranges of $\Gamma(h_2)/m_2$ allowed by the theoretical constraints in Secs.~\ref{sec:minimization} and~\ref{sec:positivedefinite} as a function of $m_2$ for $b_4=4.2$ and $\sin^2\theta=0.12$.}
\end{figure}

In Fig.~\ref{fig:totwidth} we show allowed ranges of $\Gamma(h_2)/m_2$ as a function of the mass of $m_2$ for $b_4=4.2$ and $\sin^2\theta=0.12$.  The total width is always bounded by $\Gamma(h_2)/m_2\lesssim 0.09$.  For $m_2\lesssim 700\textrm{ GeV}$, we also have $\Gamma(h_2)/m_2\lesssim 0.05$.  As $\sin\theta$ decreases below its upper bound, the total width of $h_2$ will decrease as well.  The value of $b_4$ has no effect on the partial widths of $h_2$ into SM fermions or gauge bosons.  However, as $b_4$ decreases, the partial width of $\Gamma(h_2\rightarrow h_1h_1)$ decreases as shown in Fig.~\ref{fig:paramlargeb4}.  Hence, the upper bound on $\Gamma(h_2)$ in Fig.~\ref{fig:totwidth} is the upper bound throughout the allowed parameter regions, and $h_2$ is sufficiently narrow to justify the narrow width approximation in Eq.~(\ref{eq:res}).

%%%%%%%%%%%%%%%%%%%%%%%%%%%%%%%%%%%%%%%%%%%%%%%%%%%%%%%%%%%%%
\section{\label{sec:results} Results}
%%%%%%%%%%%%%%%%%%%%%%%%%%%%%%%%%%%%%%%%%%%%%%%%%%%%%%%%%%%%%

We maximize the production rate in Eq.~(\ref{eq:res}) by fixing $m_2$ and $\theta$, then scanning over the remaining parameters
 \begin{equation}
 a_2\textrm{, } b_3\textrm{, and }b_4.
 \end{equation}
For all numerical results,  the SM production cross sections and widths for a Higgs boson in Eqs.~(\ref{eq:h1prod}), (\ref{eq:h1dec}), (\ref{eq:h2prod}), and (\ref{eq:h2SMdec}) were obtained from Ref.~\cite{deFlorian:2016spz}.

\begin{figure}[t]
\includegraphics[width=0.48\textwidth,clip]{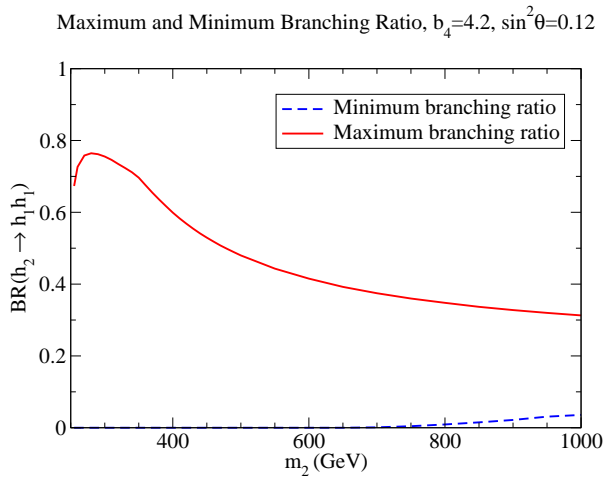}
\caption{\label{fig:minmax}Maximum and minimum allowed $\textrm{BR}(h_2 \rightarrow h_1 h_1)$ as a function of $m_2$ for $b_4=4.2$ and $\sin^2\theta=0.12$.}
\end{figure}

The maximum and minimum ${\rm BR}(h_2\rightarrow h_1h_1)$ for different values of $m_2$ are shown in Fig.~\ref{fig:minmax}.  We set $b_4=4.2$ at the perturbative unitarity bound and $\sin^2\theta=0.12$ at the experimental bound~\cite{Aad:2015pla}. The largest possible branching ratio occurs at around $280 \textrm{ GeV}$ with $\textrm{BR}(h_2 \rightarrow h_1 h_1) = 0.76$. Even up to masses of $1000 \textrm{ GeV}$ the branching ratio to double Higgs can be larger than $0.3$.  Additionally, for $m_2\gtrsim 600$~GeV there is a minimum on ${\rm BR}(h_2\rightarrow h_1 h_1)$.

\begin{figure}[t]
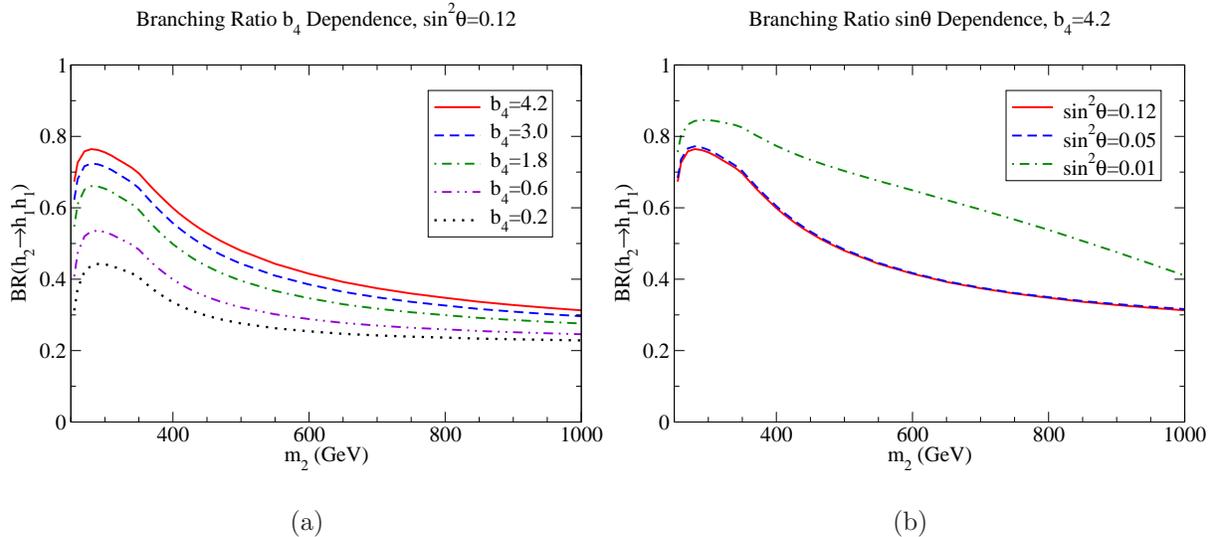

\subfigure[]{\includegraphics[width=0.48\textwidth,clip]{b4}\label{fig:b4dep}}
\subfigure[]{\includegraphics[width=0.48\textwidth,clip]{theta}\label{fig:thetadep}}
\caption{Maximum allowed $\textrm{BR}(h_2 \rightarrow h_1 h_1)$ as a function of $m_2$ for different values of (a) $b_4$ and (b) $\sin\theta$.}
\end{figure}

Figure~\ref{fig:b4dep} shows the dependence of the maximum branching ratio ${\rm BR}(h_2\rightarrow h_1h_1)$ on the parameter $b_4$.  As can be seen, if the parameter $b_4$ is less than the unitarity bound of $4.2$ then the largest possible branching ratio becomes smaller. This is due to the shrinking of the allowed range for the parameters $a_2$ and $b_3$, as shown in Fig.~\ref{fig:paramlargeb4}. Even for small values of $b_4$, the branching ratio can still be quite substantial.

The maximum possible value of $\sin^2{\theta}$ is expected to decrease as more data is taken at the LHC and the measurements of the observed Higgs couplings become more precise. Figure~\ref{fig:thetadep} shows the maximum possible ${\rm BR}(h_2\rightarrow h_1h_1)$ for several values of $\sin^2{\theta}$. As can be seen, the branching ratio can be larger for smaller $\sin\theta$.  Hence, maximization of ${\rm BR}(h_2\rightarrow h_1h_1)$ occurs at small $\sin\theta$. However, double Higgs production is not maximized with this condition.

\begin{figure}[t]
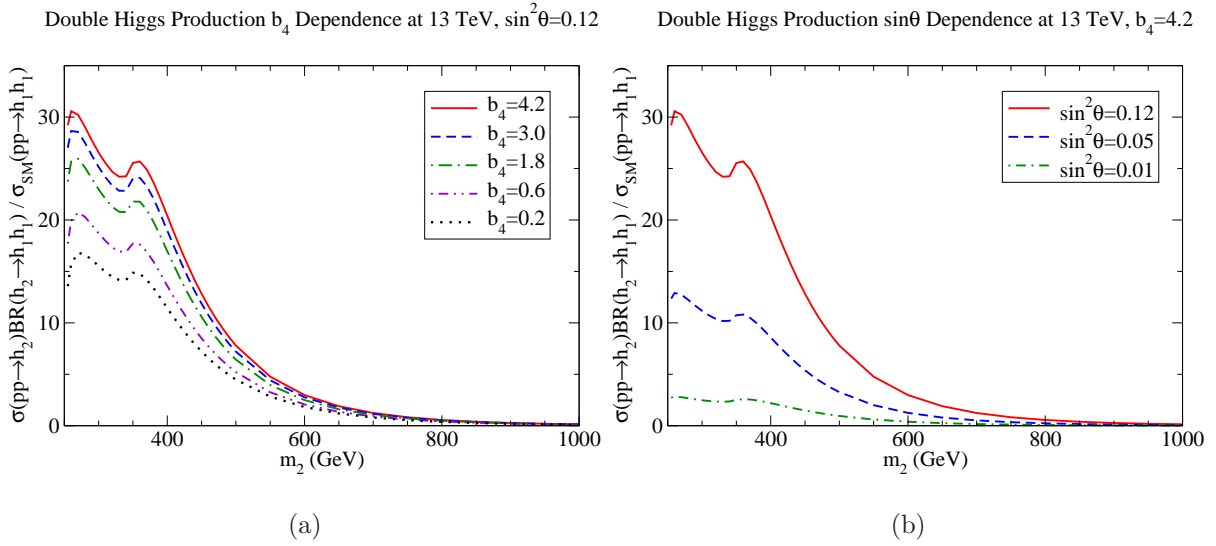

\subfigure[]{\includegraphics[width=0.48\textwidth,clip]{b4_production}\label{fig:b4prod}}
\subfigure[]{\includegraphics[width=0.48\textwidth,clip]{theta_production}\label{fig:thetaprod}}
\caption{\label{fig:prod}Maximum $\sigma(pp \rightarrow h_2) \textrm{BR}(h_2 \rightarrow h_1 h_1)$, scaled by the calculated SM double Higgs production, as a function of $m_2$ for different values of (a) $b_4$ and (b) $\sin{\theta}$.}
\end{figure}

Now we turn our attention to maximizing the double Higgs production rate. Figure~\ref{fig:prod} shows the maximum $\sigma(pp \rightarrow h_2) \textrm{BR}(h_2 \rightarrow h_1 h_1)$ at an LHC energy of $\sqrt{S_H}=13\textrm{ TeV}$ for various (a) $b_4$ and (b) $\sin \theta$ values as a function of mass $m_2$. The values are scaled by the SM double Higgs production cross section at $13\textrm{ TeV}$ of $33.53^{+5.3\%}_{-6.8\%}\textrm{ fb}$~\cite{deFlorian:2016spz}, calculated at NNLL matched to NNLO in QCD with NLO top quark mass dependence~\cite{Borowka:2016ypz}.   As mentioned earlier, the maximum rates occur when $b_4$ is at the unitarity bound $b_4=4.2$.  For $\sin\theta$, although the maximum ${\rm BR}(h_2\rightarrow h_1 h_1)$ increases as $\sin\theta$ decreases, this increase is not enough to compensate for the $\sin^2\theta$ suppression of the production cross section $\sigma(pp\rightarrow h_2)$ in Eq.~(\ref{eq:h2prod}). Hence, the maximum double Higgs production cross section occurs at the experimental bound $\sin^2\theta=0.12$.  In the best case, the resonant double Higgs production is roughly 30 times the SM double Higgs cross section.

\begin{table}
\begin{center}
\begin{tabular}{lcccc}\hline\hline
$m_2$\quad\quad\quad\quad\quad & \quad$a_2 $\quad & \quad\quad$b_3/{v_{EW}} $\quad\quad& $\quad{\rm BR}(h_2\rightarrow h_1h_2)$\quad & \quad$\sigma(pp\rightarrow h_2){\rm BR}(h_2\rightarrow h_1 h_1)$\quad\\\hline\hline
$300$~GeV & $-0.79$& $-2.7$& $0.76$& $0.89$~pb\\\hline
$400$~GeV & $-0.40$& $-3.9$& $0.60$& $0.68$~pb\\\hline
$500$~GeV & $0.059$& $-5.4$& $0.48$& $0.26$~pb\\\hline
$600$~GeV & $0.56$& $-7.1$& $0.42$& $0.10$~pb\\\hline
$700$~GeV & $1.0$& $-8.7$& $0.37$& $0.042$~pb\\\hline
$800$~GeV & $1.6$& $-11$& $0.35$& $0.019$~pb\\\hline\hline
  \end{tabular}
 \caption{\label{tab:benchmark1} Benchmark points that maximize ${\rm BR}(h_2\rightarrow h_1h_1)$ with $b_4=4.2$ and $\sin^2\theta=0.12$.   The cross sections are evaluated at a lab frame energy of $\sqrt{S_H}=13$~TeV.} \end{center}
 \end{table}

\begin{table}
\begin{center}
\begin{tabular}{lcccc}\hline\hline
$m_2$\quad\quad\quad\quad\quad & \quad$a_2 $\quad & \quad\quad$b_3/{v_{EW}} $\quad\quad& $\quad{\rm BR}(h_2\rightarrow h_1h_2)$\quad & \quad$\sigma(pp\rightarrow h_2){\rm BR}(h_2\rightarrow h_1 h_1)$\quad\\\hline\hline
$300$~GeV\ & $-1.2$& $-1.6$& $0.76$& $0.37$~pb\\\hline
$400$~GeV & $-1.0$& $-2.7$& $0.60$& $0.29$~pb\\\hline
$500$~GeV & $-0.78$& $-3.9$& $0.48$& $0.11$~pb\\\hline
$600$~GeV & $-0.59$& $-5.0$& $0.42$& $0.042$~pb\\\hline
$700$~GeV & $-0.31$& $-6.5$& $0.38$& $0.017$~pb\\\hline
$800$~GeV & $-0.015$& $-8.1$& $0.35$& $0.0079$~pb\\\hline\hline
  \end{tabular}
 \caption{\label{tab:benchmark2} Benchmark points that maximize ${\rm BR}(h_2\rightarrow h_1h_1)$ with $b_4=4.2$ and $\sin^2\theta=0.05$.   The cross sections are evaluated at a lab frame energy of $\sqrt{S_H}=13$~TeV.} \end{center}
 \end{table}

Finally, we provide our benchmark points in Tables~\ref{tab:benchmark1} and~\ref{tab:benchmark2}.  We provide the parameter points that maximize the $h_1h_1$ production in the singlet extended SM, as well as the corresponding ${\rm BR}(h_2\rightarrow h_1h_1)$ and $h_1h_1$ production cross section at a lab frame energy of $\sqrt{S_H}=13$~TeV.  As discussed before, the maximum ${\rm BR}(h_2\rightarrow h_1h_1)$ occurs for $b_4=4.2$ at the unitarity bound.  Hence, we fix $b_4=4.2$ for all benchmark points.  Also, the maximum $h_1h_1$ production cross section occurs for $\sin^2\theta=0.12$ at the current limit~\cite{Aad:2015pla}.  Table~\ref{tab:benchmark1} contains the benchmark points for $\sin^2\theta=0.12$.  However, as mentioned earlier, as the LHC continues to gather data it is expected that the precision Higgs measurements will further limit $\sin\theta$.  The uncertainties in Higgs coupling measurements are projected to be $\sim 5\%$ with $3000\textrm{ fb}^{-1}$ of integrated luminosity at the LHC~\cite{Dawson:2013bba}.  This corresponds to a bound of $\sin^2\theta \lesssim 0.05$ due to the overall $\cos^2\theta$ suppression of the $h_1$ rate of production.  Hence, we also provide benchmark points for $\sin^2\theta=0.05$ in Table~\ref{tab:benchmark2}.
%%%%%%%%%%%%%%%%%%%%%%%%%%%%%%%%%%%%%%%%%%%%%%%%%%%%%%%%%%%%%
\section{\label{sec:conc} Conclusion}
%%%%%%%%%%%%%%%%%%%%%%%%%%%%%%%%%%%%%%%%%%%%%%%%%%%%%%%%%%%%%
The simplest possible extension of the SM is the addition of a real gauge singlet scalar.  Although simple, this model is theoretically well-motivated and has interesting phenomenology.  In particular, if the new scalar $h_2$ is sufficiently heavy $m_2\geq 2\,m_1$, this model can give rise to resonant double Higgs production at the LHC.  We have investigated this signature.  We determined benchmark parameter points that maximize the double Higgs production rate in this model at the $\sqrt{S_H}=13$~TeV LHC.  These benchmark points are important for gauging when the ongoing experimental searches for resonant double Higgs production are probing interesting regions of parameter space of well-motivated models.  We have found that ${\rm BR}(h_2\rightarrow h_1 h_1)$ as high as $0.76$ and $h_1h_1$ production rates up to 30 times the SM rate are still possible.

\section*{Acknowledgements}
We thank S. Dawson for valuable discussions about double Higgs production in the singlet extended SM.  This investigation was
supported in part by the University of Kansas General Research Fund allocation 2302091.

\appendix*
\section{$v = 0$ Extrema}
In this appendix we give solutions to Eq.~(\ref{cubic}) where $v=0$. We update the results from Ref.~\cite{Chen:2014ask}, presenting these solutions in a manifestly real form.

Solutions to the extrema conditions were solutions to Eq.~(\ref{cubic}), repeated below.
\begin{equation}\label{appcubic}
b_1 + b_2 x + b_3 x^2 + b_4 x^3 = 0
\end{equation}
All the coefficients in Eq.~(\ref{appcubic}) are real. We divide Eq.~(\ref{appcubic}) by $b_4$ to normalize the cubic term:
\begin{eqnarray}
x^3 + A x^2 + B x + C = 0 \label{scaledcubic}\\
A = \frac{b_3}{b_4} \nonumber \\
B = \frac{b_2}{b_4} \nonumber \\
C = \frac{b_1}{b_4} \nonumber
\end{eqnarray}
We define the intermediate variables $Q$ and $R$ as
\begin{eqnarray}
Q = \frac{3 B - A^2}{9} \\
R = \frac{9 A B - 27 C - 2 A^3}{54}
\end{eqnarray}
The polynomial discriminant of Eq.~(\ref{scaledcubic}) is then given by
\begin{equation}
D = Q^3 + R^2
\end{equation}
The discriminant $D$ can be either positive, negative, or zero. If the discriminant is zero, the cubic has degenerate solutions. The parameter space where $D=0$ has zero volume, so it is unlikely to occur. The degenerate solutions are not important to consider for our purposes. If $D<0$, the cubic has three distinct real roots. If $D>0$, the cubic has a real root, and a pair of complex conjugate roots.

For the case $D<0$, we define an angle $\theta$ as follows:
\begin{equation}
\theta = \cos^{-1}\left( \frac{R}{-Q} \sqrt{\frac{1}{-Q}} \right)
\end{equation}
Note that if $D<0$, then we also must have $Q<0$. The three real solutions to Eq.~(\ref{appcubic}) are then given by
\begin{eqnarray}
\begin{aligned}
x_1 &= 2 \sqrt{-Q} \cos{\left( \frac{\theta}{ 3} \right) } - \frac{A}{3} \\
x_2 &= 2 \sqrt{-Q} \cos{\left( \frac{\theta + 2 \pi}{ 3} \right) } - \frac{A}{3} \\
x_3 &= 2 \sqrt{-Q} \cos{\left( \frac{\theta + 4 \pi}{ 3} \right) } - \frac{A}{3}
\end{aligned}
\end{eqnarray}

For the case $D>0$, we must look at two sub-cases. If $Q < 0$, we then define a hyperbolic angle $\eta$ as follows:
\begin{equation}
\eta = \cosh^{-1}\left( \frac{|R|}{Q} \sqrt{\frac{1}{-Q}} \right)
\end{equation}
The single real solution to Eq.~(\ref{appcubic}) is then given by
\begin{equation}
x = 2 \frac{|R|}{R} \sqrt{-Q} \cosh{\left( \frac{\eta}{3} \right) } - \frac{A}{3}
\end{equation}

For the case $D > 0$ and $Q > 0$, we also define a hyperbolic angle $\eta$:
\begin{equation}
\eta = \sinh^{-1} \left( \frac{R}{Q} \sqrt{\frac{1}{Q}} \right)
\end{equation}
The single real solution to Eq.~(\ref{appcubic}) is then given by
\begin{equation}
x = 2 \sqrt{Q} \sinh{\left( \frac{\eta}{3} \right) } - \frac{A}{3}
\end{equation}

\bibliographystyle{myutphys}
\bibliography{benchmark}
\end{document}